# A Hybrid Prefetch Scheduling Heuristic to Minimize at Run-Time the Reconfiguration Overhead of Dynamically Reconfigurable Hardware*

Javier Resano[1] (javier1@dacya.ucm.es), Daniel Mozos[1], Francky Catthoor[2]
[1] Department of Computer Architecture (DACYA), Universidad Complutense de Madrid
[2] IMEC vzw, Leuven, also Professor at Katholieke Universiteit Leuven, Belgium

**Abstract**

Due to the emergence of highly dynamic multimedia applications there is a need for flexible platforms and run-time scheduling support for embedded systems. Dynamic Reconfigurable Hardware (DRHW) is a promising candidate to provide this flexibility but, currently, not sufficient run-time scheduling support to deal with the run-time reconfigurations exists. Moreover, executing at run-time a complex scheduling heuristic to provide this support may generate an excessive run-time penalty. Hence, we have developed a hybrid design/run-time prefetch heuristic that schedules the reconfigurations at run-time, but carries out the scheduling computations at design-time by carefully identifying a set of near-optimal schedules that can be selected at run-time. This approach provides run-time flexibility with a negligible penalty.

## 1. Introduction

Current multimedia applications, such as digital video and 3D games, present highly dynamic and non-deterministic behavior, and a very variable workload. Dealing with this kind of applications involves a complex trade-off between carrying out the scheduling computation at design-time or at run-time. On the one hand, performing the whole scheduling process at design-time is very ineffective because the scheduler does not have enough information and must often assume a pessimistic worst-case scenario. On the other hand, very stringent timing requirements exist at run-time. Hence the scheduler must accomplish its task in a greatly limited time-slot applying only simple scheduling policies.

Hybrid design/run-time scheduling approaches are a very effective way to overcome this problem. They split the computation into a design-time phase and a run-time phase. The design-time phase generates sets of optimal (or near-optimal) schedules for certain run-time conditions. Later, a run-time scheduler analyses the running tasks and the run-time conditions and selects the most convenient schedule among them. The hybrid approach provides run-time flexibility and, at the same time, it generates only a small run-time penalty due to scheduler execution because most of the exploration and computation is done at design time. A very good example of this approach is the TCM [9] (Task Concurrency Management) scheduling environment initially developed for heterogeneous multiprocessor platforms. However, in order to cope with the demanding requirements of current multimedia applications, it is very interesting to extend the hybrid scheduling approach to emerging platforms containing also Dynamically Reconfigurable Hardware (DRHW) resources. These resources provide both high performance and run-time flexibility because their functionality can be updated at run-time to meet the variable requirements of the running applications. In particular, for embedded systems the amount of resources is highly constrained and, at the same time, the number of applications that they have to support is constantly increasing. In order to meet the performance requirements of these applications, specific HW support is often required. However, it is infeasible to provide Application Specific Integrated Circuits (ASICs) for all them. Using the partial reconfiguration capabilities, DRHW resources can be shared to provide this HW support for the whole set of applications.

We are targeting heterogeneous multiprocessor platforms where some of the processing elements are DRHW resources which are equivalent to any other processing elements. Hence, the scheduler assigns tasks to them at run-time according to the computational load of the system and its real-time constraints. An example of such a platform was presented in [4, 5]. Using an InterConnection Network (ICN) model (Figure 1) that provides inter-task communication and run-time allocation support, an FPGA is turned into a network-on-chip multiprocessor platform. The basic idea of the ICN model is that the DRHW resources are split into a set of identical tiles. Each tile is wrapped by a communication interface. These tiles are independently reconfigured at run-time and can communicate with each other using message-passing primitives over a network-on-a-chip. The model includes also support for embedded Instruction

---

* Research supported by the Spanish Government TIC 2002-00160

Set Processors (ISPs) and can be extended to support other types of DRHW resources, like coarse-grain arrays. This model has been successfully implemented on Virtex, and Virtex-II FPGAs [11] coupled with an ISP.

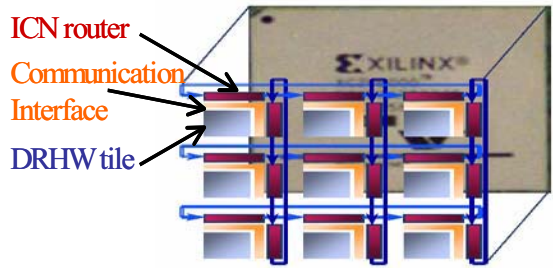

**Figure 1.** ICN model for DRHW.

As it is shown in figure 1, with the ICN model an FPGA-based platform can be considered as a multiprocessor system where subtasks are assigned to FPGA tiles instead of to ISPs. On top of this model, a multiprocessor scheduler (like TCM) can be easily applied. However, the run-time flexibility of DRHW often comes at the price of a very large reconfiguration overhead. For instance, reconfiguring one tenth of a Virtex XC2V6000 FPGA requires at least 4 ms. This overhead is not always acceptable for highly dynamic applications, since they may demand reconfigurations every few milliseconds. However, multiprocessor schedulers for embedded systems often neglect this overhead. Hence, in order to efficiently include DRHW resources, the scheduling flow must be extended adding specific support to deal with the reconfiguration overhead.

As it was explained in [6], there are two key-factors to reduce this overhead. Firstly, previously loaded subtasks must be reused. Thus, if a subtask is executed several times in a DRHW resource, it may remain loaded from one execution to another and no reconfiguration is needed. Secondly, reconfigurations must be scheduled in order to hide their latency. However, for highly dynamic applications, the reusable subtasks cannot be identified at design-time. Hence, which subtasks must be loaded and which ones can be reused is only known at run-time. Therefore, the reconfiguration schedule must be at least partially accomplished at run-time. In [7] we presented a reconfiguration scheduling technique fully performed at run-time. This technique was able to drastically reduce the reconfiguration overhead even for highly dynamic applications. However, it was not fully scalable. Hence, for large number of reconfigurations, it consumed significant time to carry out the schedule.

In this paper we present a novel hybrid design-time/run-time configuration scheduling approach that achieves almost as good results as the previous run-time heuristic while generating a very limited run-time penalty since it carries out all the computation intensive parts of the scheduling heuristic at design-time and just some minor computations are performed at run-time to tackle the non-deterministic dynamic behavior.

The remainder of the paper is organized as follows. The next section introduces the related work. Section 3 explains the reconfiguration-scheduling problem. Section 4 motivates the need of a hybrid scheduling heuristic. Sections 5 and 6 describe the prefetch scheduling design-time and run-time phases. Section 7 presents the experimental results and, finally, section 8 summarizes our conclusions.

## 2. Related work

Previously, other research groups have addressed the minimization of the reconfiguration overhead. Much of these works propose the development of new types of architectures, like multi-context FPGAs and especially coarse-grain architectures. Thus, several interesting coarse-grain platforms that can be reconfigured much faster than standard fine-grain architectures have emerged recently [1,2]. Nevertheless, the reconfigurable market is still being clearly dominated by the FPGAs.

In [12] a very interesting configuration prefetching approach to reduce the reconfiguration overhead for FPGAs is presented. This technique attempts to predict which task is going to be executed next and load it in advance. If the prediction is a success, the reconfiguration latency is, at least, partially hidden. Otherwise, an erroneous configuration is loaded with the consequent penalization. Our prefetching approach presents three main advantages compared to this one. First, it allows reducing the computational overhead, since all the prefetch decisions for a whole graph are taken at once and almost all the computation is done at design-time. Second, it prevents prediction misses, since our heuristic collaborates with a run-time scheduler receiving information about the subtasks scheduled in the near future. Finally, it reduces the overall execution time of the system, since our scheduling heuristic is aware of how its prefetch decisions affect the system performance and it uses this information to minimize the execution time.

Other good approaches regarding how to minimize the influence of the reconfiguration overhead applying scheduling techniques at design-time are found in [3] and [8]. However, they do not include any run-time component. Therefore, they can only be applied when very limited dynamic behavior exists.

## 3. Scheduling the run-time reconfigurations

In order to evaluate the hybrid prefetch technique we have developed a set of run-time modules and we have integrated them into the TCM scheduling environment [9, 10] that provides a complete framework for the experiments. Our heuristic is not specifically intended for TCM, but it can be integrated in other scheduling environments as long as they share the hybrid design-time/run-time approach.

In TCM an application is described as a set of tasks, where each task is represented as a subtask graph, that interact dynamically among them. Thus, the non-deterministic behavior must remain outside the boundaries of the tasks. If the behavior of a task depends on external data, different versions (graphs) of the same task are generated. These versions are called scenarios. In TCM the design-time scheduler generates a Pareto curve for each scenario of each task. A Pareto curve is a set of solutions where each solution is better than all the others in at least one of the parameters to optimize (in this case execution time and energy consumption). Each solution (also called Pareto point) represents an assignment and a schedule of the subtasks over the processing elements. During the execution, a run-time scheduler [10] is called periodically to identify the current scenario for each running task and select the most suitable Pareto points, i.e., those that consume less energy but still meet all the timing constraints of the application.

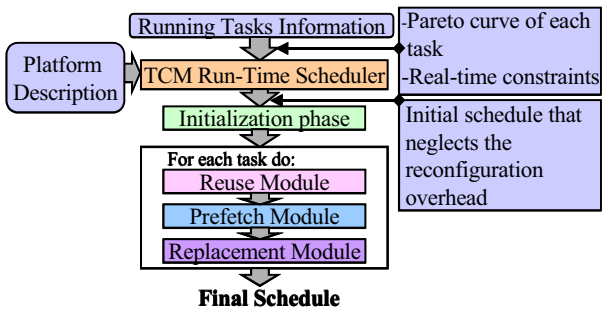

**Figure 2**. Run-time scheduling flow.

Current TCM schedulers do not take into account the reconfiguration overhead of the DRHW resources. We have provided support to tackle this specific overhead by including a set of modules in the TCM run-time scheduling flow as is depicted in Figure 2. After the run-time scheduler selects its schedule, for each task three main decisions are taken sequentially following this initial schedule. Firstly, the reuse module identifies which subtasks can be reused from a previous iteration. Secondly, if some subtasks cannot be reused, the prefetch module schedules their loads attempting to minimize the execution time overhead. Finally, when a subtask is loaded, the replacement module decides to which tile it is going to be assigned trying to maximize the percentage of reused configurations. The reuse and the replacement modules are described in detail in [6, 7]. This paper is focused on the novel hybrid prefetch module. The hybrid scheduling heuristic attempts to solve the following problem:

*Given an initial subtask schedule that neglects the reconfiguration latency, we want to update it including the needed reconfigurations scheduled in a way that minimize the overhead they generate.*

Basically, the scheduler attempts to overlap the latency of each reconfiguration with the computation of the previous subtasks. If this is possible this reconfiguration does not penalize the system performance. A simple example is depicted in Figure 3. In this figure the first schedule **(a)** is the output of a scheduler that neglects the loading overhead. The second schedule **(b)** includes the subtask loads, but does not apply any technique to reduce their impact, hence all of them introduce a delay. Finally, the third schedule **(c)** applies a configuration prefetch technique. Hence, just the first load penalizes the system performance.

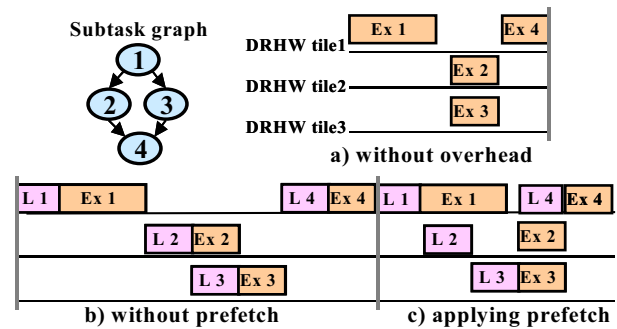

**Figure 3**. Impact of the loads over an initial subtask schedule. **L n**: load of the subtask n. **Ex n**: execution of the subtask n.

## 4. Hybrid configuration prefetch heuristic

In order to take advantage of the possibility of reusing configurations, the reconfiguration prefetch schedule must be generated at run-time. However, the time slot assigned to the run-time scheduling process is typically very small and the reconfiguration schedule is only a small part of it that must be executed many times (once for each task). Hence, it must generate good schedules very fast or it would not be applicable at run-time.

In [7] we presented a run-time reconfiguration scheduling heuristic based in list scheduling. This heuristic generated near optimal schedules, and was able to schedule 20 tasks with 14 subtasks on average in less than 0.1ms. This heuristic was developed targeting FPGAs. Therefore, the overhead generated due to the scheduling process was very small compared to the reconfiguration latency. However, currently it is also possible to include DRHW resources with much smaller reconfiguration overhead (like coarse-grain arrays). In these resources the reconfiguration overhead is still significant, but not as large as in fine grain. In addition, since the reconfiguration overhead of these architectures is more affordable, subtasks with less execution time can be assigned to them. Hence it is likely that the granularity of the subtasks decreases and, as a consequence, the number of subtasks assigned to DRHW increases. However, the complexity of our previous full run-time approach was N*Log(N), where N is the number of subtasks that must be loaded. Therefore the time needed to generate the schedule increases with the number of subtasks. For instance, increasing the size of the subtask graph by a factor of 32 was leading to a 192-increase factor in the scheduling execution time. Hence, this initial prefetch module was not fully scalable. For this reason, we have developed a new hybrid prefetch heuristic aiming to keep the good results obtained by the previous one and, at the same time, to generate almost no run-time overhead. In this heuristic the computations are split between a design-time phase and a run-time phase. Thus, all the computational intensive parts have been moved to design-time (therefore they do not generate any run-time overhead) and just some minor parts are still executed at run-time. The basic idea of this heuristic is that the design-time phase generates an optimal schedule of the reconfigurations under certain assumptions for all the possible subtask schedules selected by the TCM design-time scheduler. Later, when one of them is executed the run-time phase will guarantee that the initial assumption is true before starting its execution.

## 5. Design-time phase

An efficient prefetch technique may succeed hiding most of the reconfigurations (in [7] assuming that there was no reuse, which is the worst possible case, our heuristic was able to hide at least 75% of them). But for certain subtasks, it may fail meeting its objective because there is not always enough available time to schedule all the loads in advance (e.g. subtask 1 of Figure 3). The objective of the design-time phase of the hybrid heuristic is to identify which are those subtasks whose loading latency cannot be hidden. The hybrid heuristic is based on the definition of a subset of Critical Subtasks (CS). We define the CS subset for a given subtask graph that has been scheduled neglecting the reconfiguration overhead and a given scheduling heuristic that attempts to reduce this overhead, as the minimal subset of subtasks of the graph assigned to DRHW that fulfills the following property:

*If all the subtasks that belong to the CS subset can be reused, whereas all the remaining subtasks must be loaded, the scheduling heuristic will totally hide the latency of these loads. And therefore, they will not generate any time overhead.*

This definition is valid for any scheduling heuristic that attempts to hide the reconfiguration overhead. In our case we apply a branch&bound algorithm that always finds the optimal solution and for large graphs we keep the heuristic presented in [7] since it generates near optimal schedules in an affordable time.

Figure 4 depicts the steps followed to identify the critical subtasks of a graph. The process starts executing the prefetch scheduling heuristic assuming that none of the subtasks assigned to DRHW can be reused (hence, all of them must be loaded). Afterwards, all the subtasks that generate any delay due to its reconfiguration are detected and the one with greatest weight is included in the CS subset. These weights represent how critical is the execution of each subtask. They are assigned computing the longest path (in terms of execution time) from the beginning of the execution of the subtask to the end of the execution of the whole graph with an As-Late-As-Possible (ALAP) schedule. Hence the subtasks in the critical path always have greater weight than the others. The process continues assuming that all the subtasks assigned to the CS subset are reused until the prefetch heuristic hides the reconfiguration latencies of all the remaining subtasks assigned to DRHW. When this process finishes, the last schedule computed by the prefetch heuristic is stored. This schedule is the input of the run-time phase. In this schedule it is assumed that all the subtasks from the CS subset are reused, whereas the remaining subtasks assigned to DRHW must be loaded. This assumption means, by definition of the CS subset, that this schedule hides the latency of all these loads. Hence the reconfiguration overhead is 0.

```
For each schedule do
  1. CS := ∅;
  2. While (compute_penalty(CS) ≠ 0) do
     S:= subtasks that generate delays;
     S1:= MAX_weight(S);
     Add_subtask(S1, CS);
```

**Figure 4**. Pseudo code for the critical subtasks selection. *compute_penalty(CS)* assumes that CS subtasks are reused.

## 6. Run-time phase

The design-time schedules assume that all the nodes that belong to the CS are always loaded. However, if there are not enough DRHW resources this is not always true. The task of the run-time phase of the hybrid heuristic is to guarantee that all the subtasks from a CS subset are loaded before starting the execution of the corresponding design-time schedule. This is called initialization phase. The loading order during this phase is also decided at design-time according to the subtask weights (the subtask with the greatest weight is loaded first), hence the run-time phase must only identify which subtasks from the CS subset must be loaded.

The design-time schedule assumes that all the subtasks assigned to DRHW that do not belong to the CS subset are going to be loaded. However, if some of them can be reused it is an unnecessary waste of energy to load them again. Hence, the run-time prefetch module will cancel those loads without modifying the rest of the schedule. The only task done so far during the run-time phase of the hybrid heuristic is to identify which subtasks can be reused and which must be loaded. However, the reuse module already does this task. Hence, the prefetch hybrid heuristic is not generating any run-time overhead.

Up to now the prefetch heuristic has been always applied inside the boundaries of a task. The reason is that the actual sequence of tasks executed is only known at run-time. Therefore it is not possible to do inter-task optimizations at design-time. However, they can be performed at run-time if enough information is available. In the TCM environment the TCM run-time generates as output a sequence of scheduled tasks which can be used to apply inter-task optimizations. Using again the idea of critical subtasks, we have found a way to reduce the reconfiguration overhead introducing an inter-task optimization technique to our hybrid heuristic. Basically, for each task the run-time prefetch module uses the final idle period of the reconfiguration circuitry to carry out the initialization phase of the subsequent task. If this is possible, this task will not generate any overhead due to its reconfigurations. Figure 5 illustrates how the example introduced in Figure 3 is scheduled using the hybrid prefetch heuristic. In the picture, **b.1** is the initialization phase, where the subtask 1 (that is the only CS) is loaded. If subtask 1 could be reused b.1 would not be needed. **b.2** is the design-time schedule where the load of subtask 3 has been removed because it was reused. And **b.3** is the final time-slot when the reconfiguration circuitry was idle. This time is used to prefetch one critical subtask from the subsequent task.

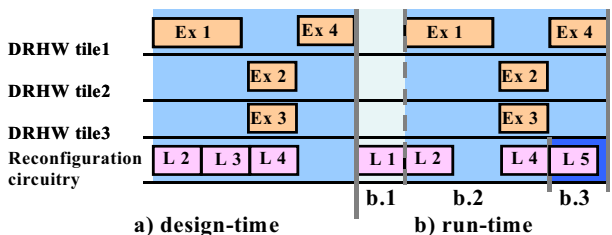

**Figure 5. a)** Schedule computed at design-time. **b)** Final schedule.

## 7. Experimental results

In order to compare our current approach with the approach presented in [7] we have applied our techniques to the same set of multimedia tasks. These tasks are a sequential and a parallel version of the JPEG decoder, an MPEG encoder, and a Pattern Recognition application that applies the Hough transform over a matrix of pixels in order to look for geometrical figures. In table 1 the features of these tasks are presented. "Ideal ex. time" is the execution time of the application when there is no reconfiguration overhead. "Overhead" is the percentage of the initial execution time that is added when the entire set of subtasks must be loaded on to the DRHW. Finally, "Prefetch" is the same overhead after applying an optimal prefetch heuristic. For the MPEG encoder there are three different scenarios corresponding to the decoding of B, P, and I frames (the table includes the average data). The appropriate scenario is selected at run-time following the sequence of frames. We have simulated 1000 iterations of the execution of this set of applications for different number of DRHW tiles assuming that the reconfiguration latency is 4 ms. In order to introduce unpredictable behavior, the applications executed during each iteration vary randomly. The simulation has been carried out five times with different prefetching approaches. The fist one did not include any prefetch module. In this case the reconfiguration overhead is 23%. In the second execution an optimal prefetch module is applied at design-time (hence it is not possible to reuse previously loaded subtasks since at design-time there is not enough information available). With this module the overhead is reduced to 7%.

**Table1.** Set of multimedia benchmarks.

| Set of Task | Sub-tasks | Ideal ex time | Overhead | Prefetch |
|---|---|---|---|---|
| Pattern Rec. | 6 | 94 ms | +17% | +4% |
| JPEG dec. | 4 | 81 ms | +20% | +5% |
| Parallel JPEG | 8 | 57 ms | +35% | +7% |
| MPEG encoder | 5 | 33 ms | +56% | +18% |

The results of the three remaining simulations are depicted in Fig 6. In this figure *run-time* are the results obtained applying the run-time heuristic from [7] with our modules that support subtask reuse. In this case, with less than 20% of the subtasks reused (for 8 tiles) the overhead is reduced to 3%. *run-time+inter-task* are the results when the run-time schedule is improved using the inter-task optimization presented in section 6 and *hybrid* are the results with the hybrid heuristic. In these two cases the overhead is at most 1.3%, hence at least 95% of the original overhead is hidden. It must be remarked that *hybrid* and *run-time+inter-task* present very similar results (of course, the run-time approach generates slightly better results). However the first approach is fully carried out at run-time, whereas the second one performs all the scheduling computations at design-time and only identifies the reusable subtasks at run-time.

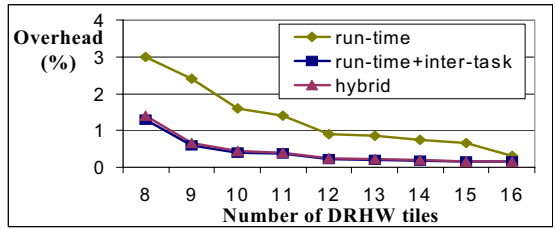

**Figure 6** Reconfiguration overhead for the 4 tasks depicted in table 1 running with dynamic behavior.

A coherent reason exists for these nice results. The hybrid heuristic generates at design-time an optimal schedule for the non-critical subtasks. Hence no run-time approach can improve this part. The critical subtasks can still generate an important overhead, but by definition, they will generate also overhead even when applying the prefetch schedule at run-time. Of course, in this case the run-time heuristic may hide it partially. However, the inter-task optimization allows hiding most of the loads of the critical subtasks. As a result our hybrid heuristic clearly outperforms the one presented in [7].

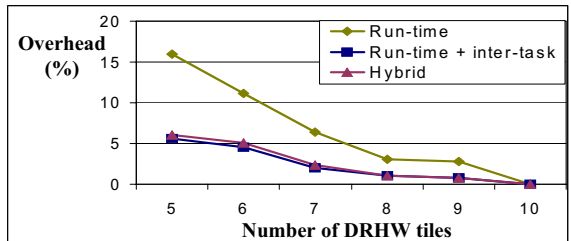

**Figure 7.** Reconfiguration overhead for a Pocket GL 3D rendering application.

We have also tested our hybrid heuristic with a highly dynamic 3D rendering application. This application is composed of 6 dynamic tasks that have in total 10 subtasks. For each task several scenarios can be selected at run-time. The amount of scenarios depends on the dynamism of the task. Thus, task 5 has four scenarios, whereas task 4 has ten. In total there are 40 different scenarios. However, due to the inter-task dependencies, at run-time just 20 feasible combinations exist, which are called inter-task scenarios. The run-time scheduler does the selection among the inter-task scenarios. The average execution time of a subtask in this application is 5.7ms, which is comparable with the 4ms needed to load a subtask onto a DRHW tile. This execution time heavily varies, going from 0.2 ms to 30ms. In this experiment 62% of the subtasks are critical. However, as it is seen in the Figure 7 the hybrid heuristic still generates almost as good results as the fully run-time approach. In this case, the reconfiguration overhead was initially 71% of the ideal execution time. Applying an optimal configuration prefetch technique at design-time it is reduced to 25%. Finally, with the hybrid heuristic the overhead is reduced to 5% for five tiles and less than 2% for eight tiles. Hence, at least 93% of the initial overhead is hidden.

## 8. Conclusions

The reconfiguration overhead of DRHW resources can drastically degrade the system performance if no active scheduling policies are applied. In addition, when dealing with dynamic applications, this problem must be tackled at run-time, when the time-slot for the scheduling process is heavily constrained. We have overcome this restriction by developing a hybrid scheduling heuristic that selects at run-time a schedule almost as good as a pure run-time approach while generating a negligible overhead since all the computation intensive parts of the scheduling process are carried out at design-time. In addition, we have improved our results by applying a simple run-time inter-task optimization technique that leads to very significant reconfiguration overhead reductions. In our experiments our hybrid heuristic has eliminated from 93% to 100% of the initial execution time overhead.

## References
[1] www.ipflex.com
[2] www.elixent.com
[3] Maestre, R. et al, "Configuration Management in Multi-Context Reconfigurable Systems", ISSS'00, pp. 107-113, 2000.
[4] Marescaux, T. et al., "Interconnection Network enable Fine-Grain Dynamic Multi-Tasking on FPGAs", FPL'02, pp. 795-805, 2002.
[5] Mignolet, J-Y. et al. "Infrastructure for Design and Management of Relocatable Tasks in a Heterogeneous Reconfigurable System-on-Chip" DATE'03, pp. 986-991, 2003.
[6] Resano, J. et al. "Specific scheduling support to minimize the reconfiguration overhead of dynamically reconfigurable hardware". DAC'04, pp. 119 – 124, 2004.
[7] Resano, J. et al. "A hybrid design-time/run-time scheduling flow to minimise the reconfiguration overhead of FPGAs". Journal on Microprocessors and Microarchitectures. Elsevier publishers. Volume 28, Issues 5-6, pp. 291-301, 2004.
[8] Shang, Li et al., "Hw/Sw Co-synthesis of Low Power Real-Time Distributed Embedded Systems with Dynamically Reconfigurable FPGAs", ASP-DAC'02, pp. 345-360, 2002.
[9] Yang, P. et al., "Energy-Aware Runtime Scheduling for Embedded-Multiprocessors SOCs", IEEE Design&Test of Computers, pp. 46-58, 2001.
[10] Yang, P. et al "Pareto-Optimization-Based Run-Time Task Scheduling for Embedded Systems". ISSS'03, pp 120-125. 2003.
[11] www.xilinx.com
[12] Zhiyuan Li, "Configuration management techniques for reconfigurable computing" Ph.D. thesis, 2002.